# A G-FDTD Method for Solving the Multi-Dimensional Time-Dependent Schrödinger Equation

*Frederick Ira Moxley III*[1] *& Weizhong Dai*[2]

1. Physics, College of Engineering & Science
   Louisiana Tech University, Ruston, LA 71272, USA
2. Mathematics & Statistics, College of Engineering & Science
   Louisiana Tech University, Ruston, LA 71272, USA

## Abstract

The Finite-Difference Time-Domain (FDTD) method is a well-known technique for the analysis of quantum devices. It solves a discretized Schrödinger equation in an explicitly iterative process. However, the method requires the spatial grid size and time step satisfy a very restricted condition in order to prevent the numerical solution from diverging. In this article, we present a generalized FDTD (G-FDTD) method for solving the multi-dimensional time-dependent linear Schrödinger equation, and obtain a more relaxed condition for stability when the finite difference approximations for spatial derivatives are employed. As such, a larger time step may be chosen. This is particularly important for quantum computations. The new G-FDTD method is tested by simulation of a particle moving in 2-D free space and then hitting an energy potential. Numerical results coincide with those obtained based on the theoretical analysis.

## 1 INTRODUCTION

The one-dimensional (1-D) time-dependent linear Schrödinger equation, which is the basis of quantum mechanics [1, 2], can be expressed as follows [3]:

$$\frac{\partial \psi(x,t)}{\partial t} = i\frac{\hbar}{2m}\frac{\partial^2 \psi(x,t)}{\partial x^2} - i\frac{V(x,t)}{\hbar}\psi(x,t), \tag{1}$$

where $m$ is the mass of the particle [kg], $\hbar = 1.054 \times 10^{-34}$ [J-sec] is Planck's constant, $V$ is the potential [J], $\psi(x,t)$ is a complex number, and $i = \sqrt{-1}$. The product of $\psi(x,t)$ with its complex conjugate, $\psi(x,t) \cdot \bar{\psi}(x,t)$, indicates the probability of a particle being at spatial location $x$ at time $t$.

It can be easily seen that the classic explicit two-level in time finite difference scheme,

*i.e.*,

$$\frac{\psi^{n+1}(k)-\psi^n(k)}{\Delta t} = \frac{\hbar}{2m\Delta x^2}\delta_x^2\psi^n(k) - i\frac{V}{\hbar}\psi^n(k), \tag{2}$$

is unconditionally unstable, where $\psi^n(k)$ is the approximation of $\psi(k\Delta x, n\Delta t)$. Here, $\Delta x$ and $\Delta t$ are the spatial grid size and time step, respectively, $k \in \mathbb{Z}$ denotes the set of all positive and negative integers, and $\delta_x^2$ is a second-order central difference operator such that

$$\delta_x^2\psi^n(k) = \psi^n(k+1) - 2\psi^n(k) + \psi^n(k-1). \tag{3}$$

Sullivan [3] applied the finite-difference time-domain (FDTD) method, which is often employed in simulations of electromagnetic fields, to solve the aforementioned Schrödinger equation. The application of the FDTD technique for the analysis of quantum devices can be described as follows [3].

The variable $\psi(x,t)$ is first split into its real and imaginary components in order to avoid using complex numbers:

$$\psi(x,t) = \psi_{real}(x,t) + i\psi_{imag}(x,t). \tag{4}$$

Inserting Eq. (4) into Eq. (1) and then separating the real and imaginary components results in the following coupled set of equations:

$$\frac{\partial\psi_{real}(x,t)}{\partial t} = -\frac{\hbar}{2m}\frac{\partial^2\psi_{imag}(x,t)}{\partial x^2} + \frac{V(x,t)}{\hbar}\psi_{imag}(x,t) \tag{5a}$$

and

$$\frac{\partial\psi_{imag}(x,t)}{\partial t} = \frac{\hbar}{2m}\frac{\partial^2\psi_{real}(x,t)}{\partial x^2} - \frac{V(x,t)}{\hbar}\psi_{real}(x,t). \tag{5b}$$

Thus, the second-order central finite difference approximations in space and time result in the FDTD scheme as follows:

$$\frac{\psi_{real}^n(k) - \psi_{real}^{n-1}(k)}{\Delta t} = -\frac{\hbar}{2m(\Delta x)^2}\delta_x^2\psi_{imag}^{n-\frac{1}{2}}(k) + \frac{1}{\hbar}V(k)\psi_{imag}^{n-\frac{1}{2}}(k) \tag{6a}$$

and

$$\frac{\psi_{imag}^{n+\frac{1}{2}}(k) - \psi_{imag}^{n-\frac{1}{2}}(k)}{\Delta t} = \frac{\hbar}{2m(\Delta x)^2}\delta_x^2\psi_{real}^n(k) - \frac{1}{\hbar}V(k)\psi_{real}^n(k). \tag{6b}$$

Here, we assume that V is dependent only on x for the sake of simplicity. The computation of the above FDTD scheme is very simple and straight-forward because one may obtain $\psi_{real}^n(k)$ from Eq. (6a) and then $\psi_{imag}^{n+1/2}(k)$ from Eq. (6b). Previously, the second author used the discrete energy method to analyze the stability of the FDTD scheme and obtained a condition for determining the time step, $\Delta t$, such that the scheme is stable [4]:

$$\frac{\hbar}{m}\cdot\frac{\Delta t}{\Delta x^2} + \frac{\Delta t}{2\hbar}\max|V| \leq c < 1, \tag{7}$$

where c is a constant. It should be pointed out that Soriano et al. [5] also used the eigenvalue method to analyze the stability of the FDTD scheme and obtained a very similar condition of $\frac{\hbar}{m} \cdot \frac{\Delta t}{\Delta x^2} + \frac{\Delta t}{2\hbar} max|V| \leq 1$. However, as pointed out in [4], even if the condition $\frac{\hbar}{m} \cdot \frac{\Delta t}{\Delta x^2} + \frac{\Delta t}{2\hbar} max|V| < 1$ is chosen, the numerical solution may still diverge. Eq. (7) indicates that the condition for stability must be less than 1 but not equal to 1.

The motivation of this study is to relax the above restriction on the mesh ratio, $\frac{\Delta t}{\Delta x^2}$, by developing a generalized FDTD (G-FDTD) scheme for solving the multi-dimensional Schrödinger equation. As such, a larger time step may be chosen. This is particularly important for quantum computations.

## 2 GENERALIZED FDTD METHOD

Recently, we have developed a G-FDTD scheme [6, 7]. The idea is that we assume $\psi_{real}(x,t)$ and $\psi_{imag}(x,t)$ are sufficiently smooth functions which vanish for sufficiently large $|x|$ and the potential $V$ is dependent only on $x$. We first rewrite Eqs. (5a) and (5b) as

$$\frac{\partial \psi_{real}(x,t)}{\partial t} = \left(-\frac{\hbar}{2m}A + \frac{V}{\hbar}\right)\psi_{imag}(x,t), \quad (8a)$$

$$\frac{\partial \psi_{imag}(x,t)}{\partial t} = \left(\frac{\hbar}{2m}A - \frac{V}{\hbar}\right)\psi_{real}(x,t), \quad (8b)$$

where $A = \frac{\partial^2}{\partial x^2}$, and employ the Taylor series method to expand $\psi_{real}(x,t_n)$ and $\psi_{real}(x,t_{n-1})$ at $t = t_{n-1/2} = (n-\frac{1}{2})\Delta t$ as follows:

$$\psi_{real}(x,t_n) = \psi_{real}(x,t_{n-1}) + 2\sum_{p=0}^{\infty}\left(\frac{\Delta t}{2}\right)^{2p+1}\frac{1}{(2p+1)!}\frac{\partial^{2p+1}\psi_{real}(x,t_{n-1/2})}{\partial t^{2p+1}}. \quad (9)$$

We then evaluate those derivatives in Eq. (9) by using Eqs. (8a) and (8b) repeatedly:

$$\frac{\partial \psi_{real}(x,t_{n-1/2})}{\partial t} = -\left(\frac{\hbar}{2m}A - \frac{V}{\hbar}\right)\psi_{imag}(x,t_{n-1/2}), \quad (10a)$$

$$\frac{\partial^3 \psi_{real}(x,t_{n-1/2})}{\partial t^3} = -\left(\frac{\hbar}{2m}A - \frac{V}{\hbar}\right)\frac{\partial^2 \psi_{imag}(x,t_{n-1/2})}{\partial t^2}$$

$$= -\left(\frac{\hbar}{2m}A - \frac{V}{\hbar}\right)\left(\frac{\hbar}{2m}A - \frac{V}{\hbar}\right)\frac{\partial \psi_{real}(x,t_{n-1/2})}{\partial t}$$

$$= \left(\frac{\hbar}{2m}A - \frac{V}{\hbar}\right)^3 \psi_{imag}(x,t_{n-1/2}), \quad (10b)$$

$$\frac{\partial^5 \psi_{real}(x,t_{n-1/2})}{\partial t^5} = \left(\frac{\hbar}{2m}A - \frac{V}{\hbar}\right)^3 \frac{\partial^2 \psi_{imag}(x,t_{n-1/2})}{\partial t^2}$$

$$= \left(\frac{\hbar}{2m}A - \frac{V}{\hbar}\right)^3 \left(\frac{\hbar}{2m}A - \frac{V}{\hbar}\right) \frac{\partial \psi_{real}(x,t_{n-1/2})}{\partial t}$$

$$= -\left(\frac{\hbar}{2m}A - \frac{V}{\hbar}\right)^5 \psi_{imag}(x,t_{n-1/2}), \tag{10c}$$

and so on. Substituting Eq. (10) into Eq. (9) gives

$$\psi_{real}(x,t_n) = \psi_{real}(x,t_{n-1}) + 2\sum_{p=0}^{N}\left(\frac{\Delta t}{2}\right)^{2p+1}\frac{(-1)^{p+1}}{(2p+1)!}\left(\frac{\hbar}{2m}A - \frac{V}{\hbar}\right)^{2p+1}\psi_{imag}(x,t_{n-1/2})$$
$$+ O(\Delta t^{2N+3}). \tag{11}$$

Similarly, we employ the Taylor series method to expand $\psi_{imag}(x, t_{n+1/2})$ and $\psi_{imag}(x, t_{n-1/2})$ at $t = t_n = n\Delta t$ as follows:

$$\psi_{imag}(x,t_{n+1/2}) = \psi_{imag}(x,t_{n-1/2}) + 2\sum_{p=0}^{N}\left(\frac{\Delta t}{2}\right)^{2p+1}\frac{1}{(2p+1)!}\frac{\partial^{2p+1}\psi_{imag}(x,t_n)}{\partial t^{2p+1}}. \tag{12}$$

Again, using Eqs. (8a) and (8b) repeatedly to evaluate those derivatives in Eq. (12), we obtain

$$\frac{\partial \psi_{imag}(x,t_n)}{\partial t} = \left(\frac{\hbar}{2m}A - \frac{V}{\hbar}\right)\psi_{real}(x,t_n), \tag{13a}$$

$$\frac{\partial^3 \psi_{imag}(x,t_n)}{\partial t^3} = \left(\frac{\hbar}{2m}A - \frac{V}{\hbar}\right)\frac{\partial^2 \psi_{real}(x,t_n)}{\partial t^2}$$
$$= -\left(\frac{\hbar}{2m}A - \frac{V}{\hbar}\right)\left(\frac{\hbar}{2m}A - \frac{V}{\hbar}\right)\frac{\partial \psi_{imag}(x,t_n)}{\partial t}$$
$$= -\left(\frac{\hbar}{2m}A - \frac{V}{\hbar}\right)^3 \psi_{real}(x,t_n), \tag{13b}$$

$$\frac{\partial^5 \psi_{imag}(x,t_n)}{\partial t^5} = -\left(\frac{\hbar}{2m}A - \frac{V}{\hbar}\right)^3 \frac{\partial^2 \psi_{real}(x,t_n)}{\partial t^2}$$
$$= \left(\frac{\hbar}{2m}A - \frac{V}{\hbar}\right)^3 \left(\frac{\hbar}{2m}A - \frac{V}{\hbar}\right)\frac{\partial \psi_{imag}(x,t_n)}{\partial t}$$
$$= \left(\frac{\hbar}{2m}A - \frac{V}{\hbar}\right)^5 \psi_{real}(x,t_n), \tag{13c}$$

and so on. Substituting Eq. (13) into Eq. (12) gives

$$\psi_{imag}(x,t_{n+1/2}) = \psi_{imag}(x,t_{n-1/2}) + 2\sum_{p=0}^{N}\left(\frac{\Delta t}{2}\right)^{2p+1}\frac{(-1)^p}{(2p+1)!}\left(\frac{\hbar}{2m}A - \frac{V}{\hbar}\right)^{2p+1}\psi_{real}(x,t_n)$$
$$+ O(\Delta t^{2N+3}). \tag{14}$$

Thus, if $A\psi_{imag}(k\Delta x, t_{n-1/2})$ and $A\psi_{real}(k\Delta x, t_n)$ are approximated using some accurate finite differences, one may obtain a G-FDTD scheme for solving the time-dependent Schrödinger equation as follows:

$$\psi_{real}^n(k) = \psi_{real}^{n-1}(k) + 2\sum_{p=0}^N \left(\frac{\Delta t}{2}\right)^{2p+1} \frac{(-1)^{p+1}}{(2p+1)!} \left(\frac{\hbar}{2m}A - \frac{V}{\hbar}\right)^{2p+1} \psi_{imag}^{n-1/2}(k), \tag{15a}$$

$$\psi_{imag}^{n+1/2}(k) = \psi_{imag}^{n-1/2}(k) + 2\sum_{p=0}^N \left(\frac{\Delta t}{2}\right)^{2p+1} \frac{(-1)^p}{(2p+1)!} \left(\frac{\hbar}{2m}A - \frac{V}{\hbar}\right)^{2p+1} \psi_{real}^n(k). \tag{15b}$$

We now extend the above G-FDTD scheme to multi-dimensional problems. For instance, we may approximate the two-dimensional (2-D) Laplace operator A by a second-order 2-D central difference operator $\frac{1}{\Delta x^2}\delta_x^2 + \frac{1}{\Delta y^2}\delta_y^2$ and a fourth-order 2-D central difference operator $\frac{1}{\Delta x^2}D_x^2 + \frac{1}{\Delta y^2}D_y^2$, respectively, where

$$\begin{aligned}
A\psi_{real}^n(j,k) &\approx \frac{1}{\Delta x^2}\delta_x^2\psi_{real}^n(j,k) + \frac{1}{\Delta y^2}\delta_y^2\psi_{real}^n(j,k) \\
&= \frac{1}{\Delta x^2}[\psi_{real}^n(j+1,k) - 2\psi_{real}^n(j,k) + \psi_{real}^n(j-1,k)] \\
&\quad + \frac{1}{\Delta y^2}[\psi_{real}^n(j,k+1) - 2\psi_{real}^n(j,k) + \psi_{real}^n(j,k-1)], \\
&\quad \text{with } O(\Delta x^2 + \Delta y^2),
\end{aligned} \tag{16a}$$

$$\begin{aligned}
A\psi_{real}^n(j,k) &\approx \frac{1}{\Delta x^2}D_x^2\psi_{real}^n(j,k) + \frac{1}{\Delta y^2}D_y^2\psi_{real}^n(j,k) \\
&= \frac{1}{12\Delta x^2}[-\psi_{real}^n(j+2,k) + 16\psi_{real}^n(j+1,k) - 30\psi_{real}^n(j,k) \\
&\quad + 16\psi_{real}^n(j-1,k) - \psi_{real}^n(j-2,k)] \\
&\quad + \frac{1}{12\Delta y^2}[-\psi_{real}^n(j,k+2) + 16\psi_{real}^n(j,k+1) - 30\psi_{real}^n(j,k) \\
&\quad + 16\psi_{real}^n(j,k-1) - \psi_{real}^n(j,k-2)], \\
&\quad \text{with } O(\Delta x^4 + \Delta y^4),
\end{aligned} \tag{16b}$$

and similar finite difference approximations for $A\psi_{imag}^{n+\frac{1}{2}}(j,k)$. We assume that $V$ is a constant and use the Von Neumann analysis [8] to analyze the stability of the G-FDTD schemes. We first let $\psi_{real}^n(j,k) = \lambda_{real}^n e^{i(j\beta_x\Delta x + k\beta_y\Delta y)}$ and $\psi_{imag}^{n+1/2}(j,k) = \lambda_{imag}^n e^{i(j\beta_x\Delta x + k\beta_y\Delta y)}$, and substitute them into Eq. (16a). This gives

$$\begin{aligned}
&\frac{1}{\Delta x^2}\delta_x^2\psi_{real}^n(j,k) + \frac{1}{\Delta y^2}\delta_y^2\psi_{real}^n(j,k) \\
&= \left[\frac{1}{\Delta x^2}\left(-4\sin^2\frac{\beta_x\Delta x}{2}\right) + \frac{1}{\Delta y^2}\left(-4\sin^2\frac{\beta_y\Delta y}{2}\right)\right]\lambda_{real}^n e^{i(j\beta_x\Delta x + k\beta_y\Delta y)},
\end{aligned} \tag{17a}$$

$$\begin{aligned}
&\frac{1}{\Delta x^2}\delta_x^2\psi_{imag}^{n+1/2}(j,k) + \frac{1}{\Delta y^2}\delta_y^2\psi_{imag}^{n+1/2}(j,k) \\
&= \left[\frac{1}{\Delta x^2}\left(-4\sin^2\frac{\beta_x\Delta x}{2}\right) + \frac{1}{\Delta y^2}\left(-4\sin^2\frac{\beta_y\Delta y}{2}\right)\right]\lambda_{imag}^n e^{i(j\beta_x\Delta x + k\beta_y\Delta y)}.
\end{aligned} \tag{17b}$$

Replacing $A$ with $\frac{1}{\Delta x^2}\delta_x^2 + \frac{1}{\Delta y^2}\delta_y^2$ in Eq. (15), substituting Eq. (17) into the resulting equations, and then deleting the common factor $e^{i(j\beta_x\Delta x + k\beta_y\Delta y)}$, we obtain

$$\lambda_{real}^n = \lambda_{real}^{n-1} + 2\sum_{p=0}^{N}\frac{(-1)^p}{(2p+1)!}[\frac{\hbar}{m}(r_x \sin^2\frac{\beta_x\Delta x}{2} + r_y \sin^2\frac{\beta_y\Delta y}{2}) + \frac{V\Delta t}{2\hbar}]^{2p+1}\lambda_{imag}^{n-1}, \quad (18a)$$

$$\lambda_{imag}^n = \lambda_{imag}^{n-1} + 2\sum_{p=0}^{N}\frac{(-1)^{p+1}}{(2p+1)!}[\frac{\hbar}{m}(r_x \sin^2\frac{\beta_x\Delta x}{2} + r_y \sin^2\frac{\beta_y\Delta y}{2}) + \frac{V\Delta t}{2\hbar}]^{2p+1}\lambda_{real}^n, \quad (18b)$$

where $r_x = \frac{\Delta t}{\Delta x^2}$ and $r_y = \frac{\Delta t}{\Delta y^2}$. Thus, we obtain a quadratic equation for $\lambda_{real}$ as follows:

$$\lambda_{real}^2 - (2 - \alpha^2)\lambda_{real} + 1 = 0, \quad (19)$$

where $\alpha = 2\sum_{p=0}^{N}\frac{(-1)^p}{(2p+1)!}[\frac{\hbar}{m}(r_x \sin^2\frac{\beta_x\Delta x}{2} + r_y \sin^2\frac{\beta_y\Delta y}{2}) + \frac{V\Delta t}{2\hbar}]^{2p+1}$. By the Von Neumann analysis, we conclude that the G-FDTD scheme is stable if $|\alpha| \leq 2$, i.e.,

$$\left|\sum_{p=0}^{N}\frac{(-1)^p}{(2p+1)!}[\frac{\hbar}{m}(r_x \sin^2\frac{\beta_x\Delta x}{2} + r_y \sin^2\frac{\beta_y\Delta y}{2}) + \frac{V\Delta t}{2\hbar}]^{2p+1}\right| \leq 1. \quad (20)$$

Hence, we obtain the following theorem.

**Theorem 1.** The generalized 2-D FDTD scheme

$$\psi_{real}^n(j,k) = \psi_{real}^{n-1}(j,k) + 2\sum_{p=0}^{N}\frac{(-1)^{p+1}}{(2p+1)!}[\frac{\hbar}{4m}(r_x\delta_x^2 + r_y\delta_y^2) - \frac{V\Delta t}{2\hbar}]^{2p+1}\psi_{imag}^{n-1/2}(j,k), \quad (21a)$$

$$\psi_{imag}^{n+1/2}(j,k) = \psi_{imag}^{n-1/2}(j,k) + 2\sum_{p=0}^{N}\frac{(-1)^p}{(2p+1)!}[\frac{\hbar}{4m}(r_x\delta_x^2 + r_y\delta_y^2) - \frac{V\Delta t}{2\hbar}]^{2p+1}\psi_{real}^n(j,k). \quad (21b)$$

is stable if the condition

$$\left|\sum_{p=0}^{N}\frac{(-1)^p}{(2p+1)!}[\frac{\hbar}{m}(r_x + r_y) + \frac{V\Delta t}{2\hbar}]^{2p+1}\right| \leq c < 1, \quad (22)$$

where $c$ is a constant.

Similarly, for the fourth-order central difference case, we let $\psi_{real}^n(j,k) = \lambda_{real}^n e^{i(j\beta_x\Delta x + k\beta_y\Delta y)}$ and $\psi_{imag}^{n+1/2}(j,k) = \lambda_{imag}^n e^{i(j\beta_x\Delta x + k\beta_y\Delta y)}$ and substitute them into Eq. (16b). This gives

$$\frac{1}{\Delta x^2}D_x^2\psi_{real}^n(j,k) + \frac{1}{\Delta y^2}D_y^2\psi_{real}^n(j,k)$$
$$= -[\frac{4}{3\Delta x^2}\sin^2\frac{\beta_x\Delta x}{2}\left(3 + \sin^2\frac{\beta_x\Delta x}{2}\right) + \frac{4}{3\Delta y^2}\sin^2\frac{\beta_y\Delta y}{2}\left(3 + \sin^2\frac{\beta_y\Delta y}{2}\right)]$$
$$\cdot \lambda_{real}^n e^{i(j\beta_x\Delta x + k\beta_y\Delta y)}, \quad (23a)$$

$$\frac{1}{\Delta x^2}D_x^2\psi_{imag}^{n+1/2}(j,k) + \frac{1}{\Delta y^2}D_y^2\psi_{imag}^{n+1/2}(j,k)$$
$$= -[\frac{4}{3\Delta x^2}\sin^2\frac{\beta_x\Delta x}{2}\left(3 + \sin^2\frac{\beta_x\Delta x}{2}\right) + \frac{4}{3\Delta y^2}\sin^2\frac{\beta_y\Delta y}{2}\left(3 + \sin^2\frac{\beta_y\Delta y}{2}\right)]$$

$$\cdot \lambda_{imag}^n e^{i(j\beta_x \Delta x + k\beta_y \Delta y)}. \tag{23b}$$

Replacing $A$ with $\frac{1}{\Delta x^2}D_x^2 + \frac{1}{\Delta y^2}D_y^2$, substituting Eq. (23) into Eq. (15), and deleting the common factor $e^{i(j\beta_x \Delta x + k\beta_y \Delta y)}$, we obtain a quadratic equation for $\lambda_{real}$ as follows:

$$\lambda_{real}^2 - (2-\alpha^2)\lambda_{real} + 1 = 0, \tag{24}$$

where

$$\alpha = 2\sum_{p=0}^{N} \frac{(-1)^p}{(2p+1)!}\{\frac{\hbar}{3m}[r_x \sin^2 \frac{\beta_x \Delta x}{2}(3+\sin^2 \frac{\beta_x \Delta x}{2}) + r_y \sin^2 \frac{\beta_y \Delta y}{2}(3+\sin^2 \frac{\beta_y \Delta y}{2})] + \frac{V\Delta t}{2\hbar}\}^{2p+1}.$$

Hence, we use a similar argument as before and obtain the following theorem.

**Theorem 2.** The generalized 2-D FDTD scheme

$$\psi_{real}^n(j,k) = \psi_{real}^{n-1}(j,k) + 2\sum_{p=0}^{N}\frac{(-1)^{p+1}}{(2p+1)!}[\frac{\hbar}{4m}(r_x D_x^2 + r_y D_y^2) - \frac{V\Delta t}{2\hbar}]^{2p+1}\psi_{imag}^{n-1/2}(j,k), \tag{25a}$$

$$\psi_{imag}^{n+1/2}(j,k) = \psi_{imag}^{n-1/2}(j,k) + 2\sum_{p=0}^{N}\frac{(-1)^p}{(2p+1)!}[\frac{\hbar}{4m}(r_x D_x^2 + r_y D_y^2) - \frac{V\Delta t}{2\hbar}]^{2p+1}\psi_{real}^n(j,k). \tag{25b}$$

is stable if the following condition is satisfied

$$\left|\sum_{p=0}^{N}\frac{(-1)^p}{(2p+1)!}[\frac{4\hbar}{3m}(r_x + r_y) + \frac{V\Delta t}{2\hbar}]^{2p+1}\right| \leq c < 1, \tag{26}$$

where $c$ is a constant.

One may employ much higher-order 2-D central difference approximations for the Laplace operator $A$ and obtain the stability condition using a similar argument. Moreover, in the GFDTD schemes, such as Eq. (21), the values of $(r_x \delta_x^2 + r_y \delta_y^2)^{2p+1}$ on $\psi_{imag}^{n-1/2}(j,k)$ and $\psi_{real}^n(j,k)$ need to be calculated, and this can be very tedious for large integer $p$. Consequently, one may use Eq. (16a) (with both sides multiplied by $\Delta t$) recursively to obtain an approximation for $(r_x \delta_x^2 + r_y \delta_y^2)^{2p+1}\psi_{real}^n(j,k)$. Similarly, we may obtain an approximation for $(r_x \delta_x^2 + r_y \delta_y^2)^{2p+1}\psi_{imag}^{n-1/2}(j,k)$. Finally, one may use a similar argument to obtain three-dimensional G-FDTD schemes.

## 3 NUMERICAL EXAMPLE

To test the stability of the 2-D G-FDTD schemes in Eq. (21) and Eq. (25), we study a 2-D problem where a particle moves in 2-D free space and then hits an energy potential. To this end, we initiated a particle at a wavelength of $\lambda$ in a Gaussian envelop of width $\sigma$ with the following two equations:

$$\psi_{real}^{0}(j,k) = \exp\left(-\frac{1}{2}\left(\frac{j-j_{0}}{\sigma}\right)^{2} - \frac{1}{2}\left(\frac{k-k_{0}}{\sigma}\right)^{2}\right)\cos\left(\frac{2\pi(j-j_{0}) + 2\pi(k-k_{0})}{\lambda}\right) \quad (27a)$$

and

$$\psi_{imag}^{0}(j,k) = \exp\left(-\frac{1}{2}\left(\frac{j-j_{0}}{\sigma}\right)^{2} - \frac{1}{2}\left(\frac{k-k_{0}}{\sigma}\right)^{2}\right)\sin\left(\frac{2\pi(j-j_{0}) + 2\pi(k-k_{0})}{\lambda}\right), \quad (27b)$$

where $k_0$ is the center of the pulse. We chose a mesh of $800 \times 800$ spatial grid points and $\Delta x = \Delta y = 0.1 \times 10^{-10} [m]$, $k_0 = 200$, and $\sigma = \lambda = 1.0 \times 10^{-10}[m]$. Furthermore, $V(j,k)$ was chosen to be 100 [eV] when $j = k = 401, \ldots, 800$, and 0 [eV] in the rest of the grid points. The electron moves in 2-D free space and then hits an energy potential with a total energy of about 300 [eV]. In our computations, we chose $N = 2$ in Eq. (21) and Eq. (25), and let $\Delta t = \mu \frac{2m\Delta x^2}{\hbar}$. Hence, we may rewrite the conditions in Eq. (22) and Eq. (26) for $N = 2$ as, respectively,

$$\left|\sum_{p=0}^{2}\frac{(-1)^{p}}{(2p+1)!}\left(4\mu + \frac{\max|V|\Delta t}{2\hbar}\right)^{2p+1}\right| \leq c < 1, \quad (28a)$$

$$\left|\sum_{p=0}^{2}\frac{(-1)^{p}}{(2p+1)!}\left(\frac{16}{3}\mu + \frac{\max|V|\Delta t}{2\hbar}\right)^{2p+1}\right| \leq c < 1. \quad (28b)$$

Figures 1-2 show the simulation of an electron moving in 2-D free space and then hitting a potential of 100 [eV], which was obtained using the original 2-D FDTD scheme, Eq. (21), with $N = 0$ and μ chosen to be 0.2, and 0.25, respectively. From these two figures, one may see that when μ = 0.2 (in which $\frac{\hbar}{m}r + \frac{\Delta t}{\hbar}max|V| = 4\mu + \frac{\Delta t}{2\hbar}max|V| \leq 0.8 + 1.7 \times 10^{-26} < 1$), the 2-D FDTD scheme is stable and its numerical solution does not diverge. However, when μ = 0.25 (in which $\frac{\hbar}{m}r + \frac{\Delta t}{\hbar}max|V| = 4\mu + \frac{\Delta t}{2\hbar}max|V| > 1$), the numerical solution is divergent.

Figures 3-4 show the simulation of an electron moving in 2-D free space and then hitting a potential of 100 [eV], which was obtained using the 2-D G-FDTD scheme, Eq. (21), with $N = 2$ and μ chosen to be 0.25, and 0.35, respectively. When μ = 0.25 (in which $\sum_{p=0}^{2}\frac{(-1)^p}{(2p+1)!}(4\mu + \frac{max|V|\Delta t}{2\hbar})^{2p+1} \leq 0.8418 + 1.7 \times 10^{-26} < 1$) and μ = 0.35 (in which $\sum_{p=0}^{2}\frac{(-1)^p}{(2p+1)!}(4\mu + \frac{max|V|\Delta t}{2\hbar})^{2p+1} \leq 0.9875 + 1.7 \times 10^{-26} < 1$), the scheme provides stable solutions.

Figure 5 shows the simulation of an electron moving in 2-D free space and then hitting a potential of 100 [eV], which was obtained using the generalized 2-D FDTD scheme, Eq. (25), with $N = 2$ and μ chosen to be 0.25 (in which $\sum_{p=0}^{2}\frac{(-1)^p}{(2p+1)!}(\frac{16}{3}\mu + \frac{max|V|\Delta t}{2\hbar})^{2p+1} \leq 0.9735 + 1.7 \times 10^{-26} < 1$). It can be seen from this figure that the 2-D G-FDTD scheme gives a stable solution.

We conclude from the above numerical example that both 2-D G-FDTD schemes improve the limitation (μ = 0.25) of the original 2-D FDTD scheme.

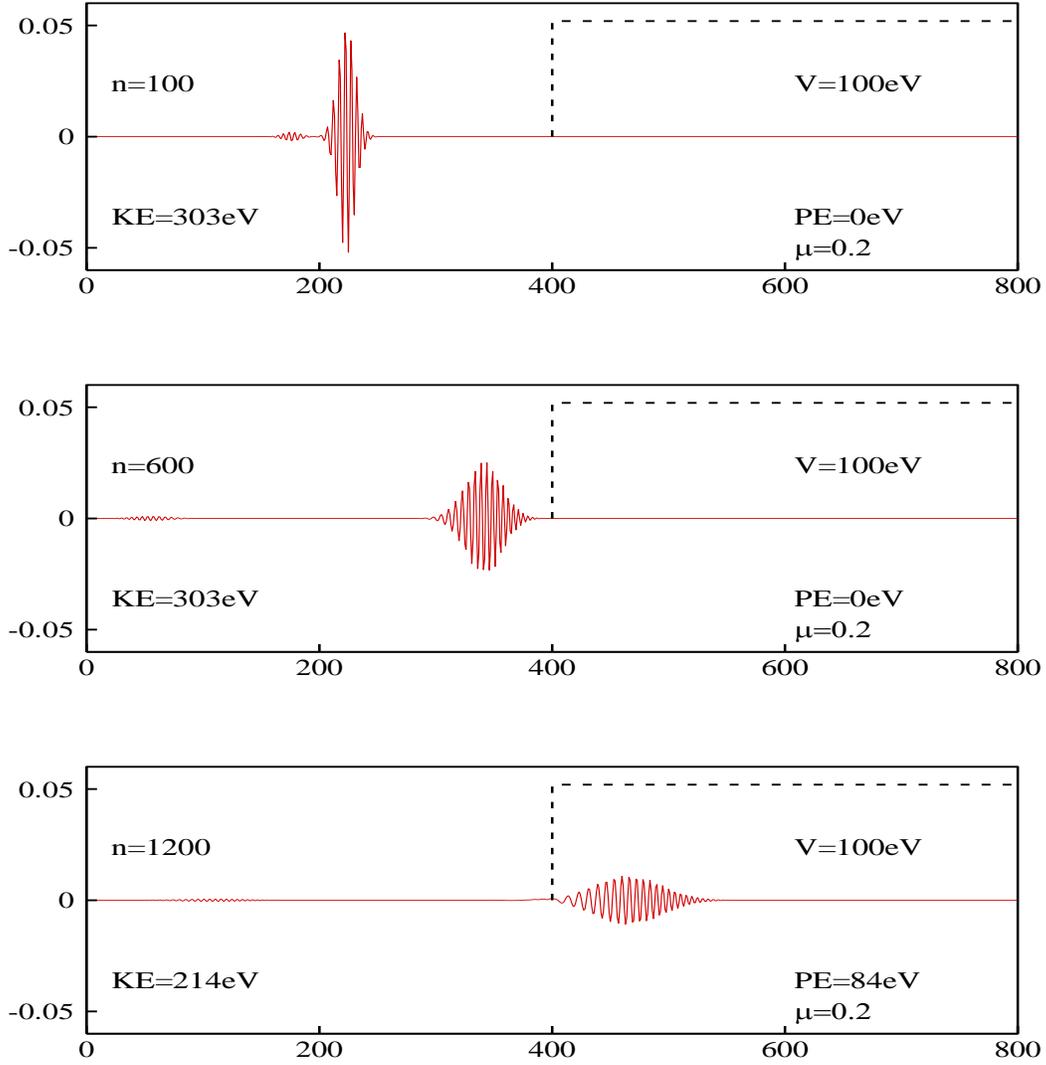

FIG. 1. Simulation of an electron moving in 2-D free space and then hitting a potential. The original 2-D FDTD scheme was employed with $\mu = 0.2$. Here, the horizontal coordinate is "$k$" and the vertical coordinate is $\psi_{real}(k,k)$.

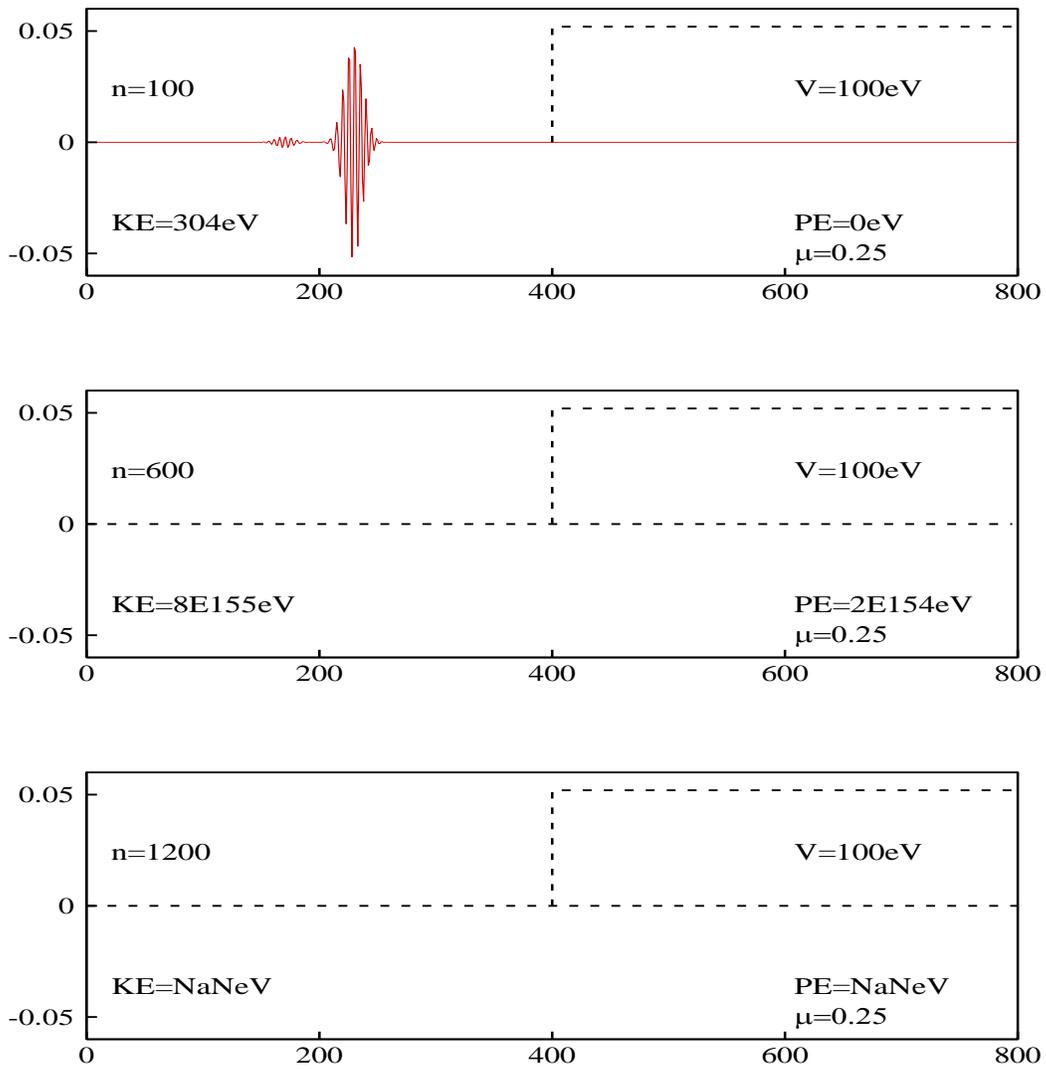

FIG. 2. Simulation of an electron moving in 2-D free space and then hitting a potential. The original 2-D FDTD scheme was employed with $\mu = 0.25$.

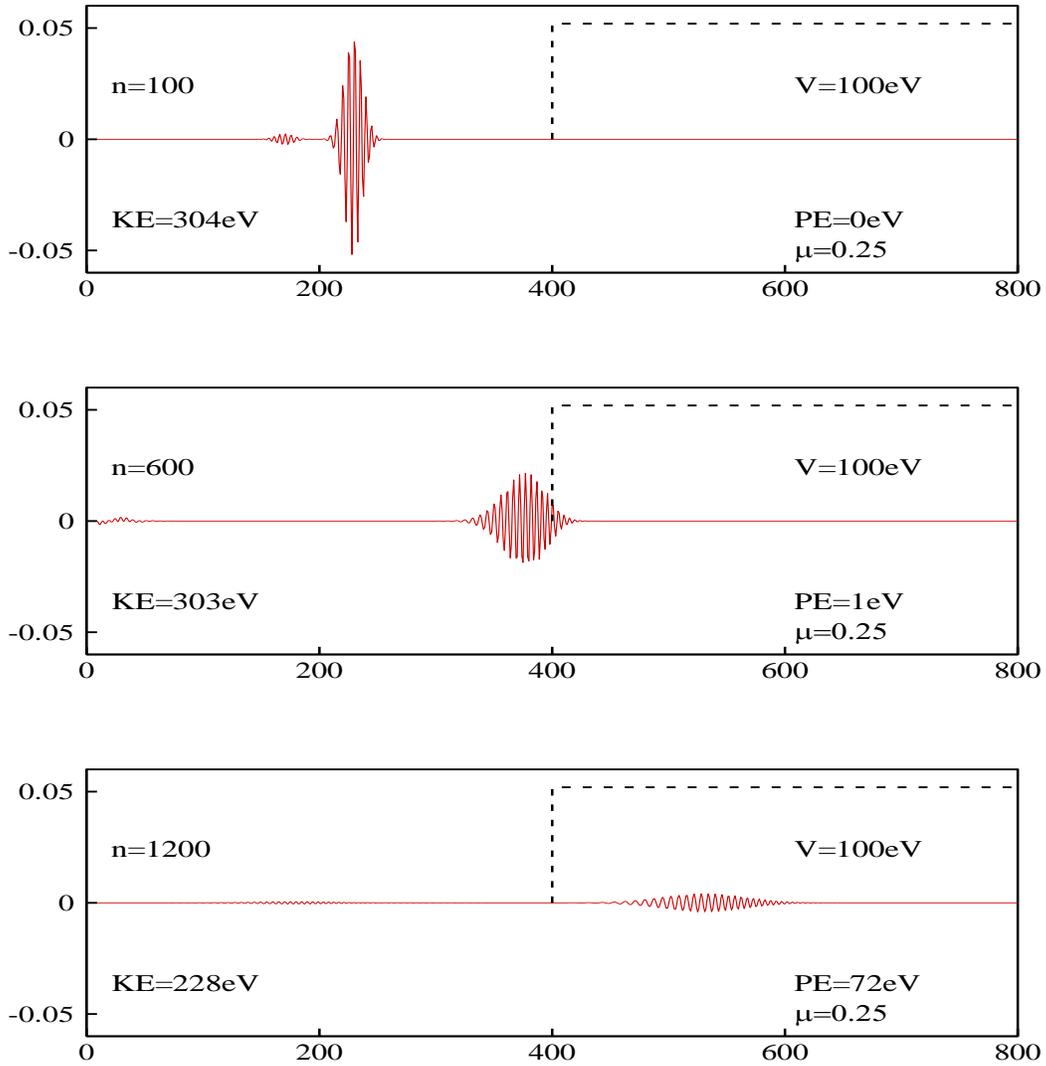

FIG. 3.  Simulation of an electron moving in 2-D free space and then hitting a potential. The generalized 2-D FDTD scheme, Eq. (34), was employed with $\mu = 0.25$.

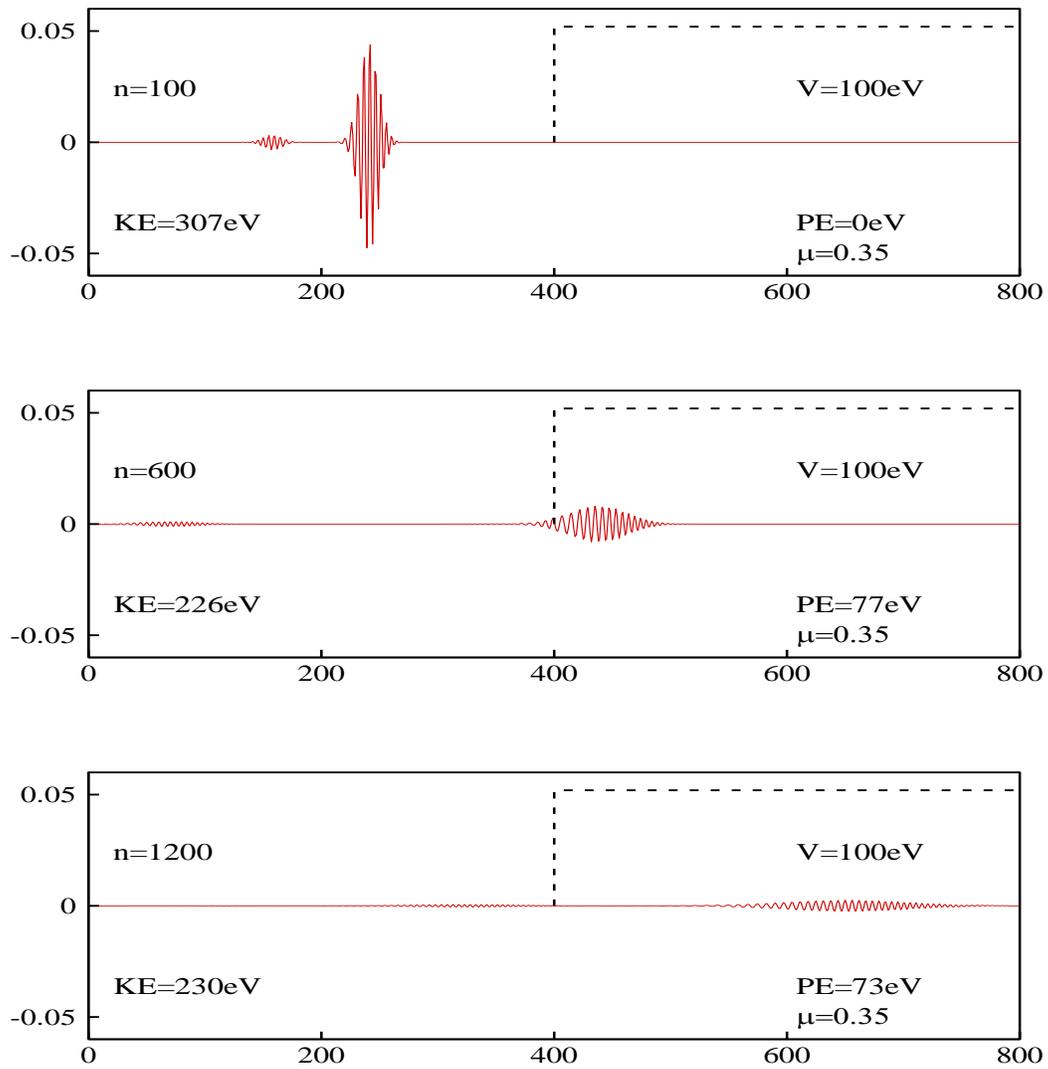

FIG. 4. Simulation of an electron moving in 2-D free space and then hitting a potential. The generalized 2-D FDTD scheme, Eq. (34), was employed with $\mu = 0.35$.

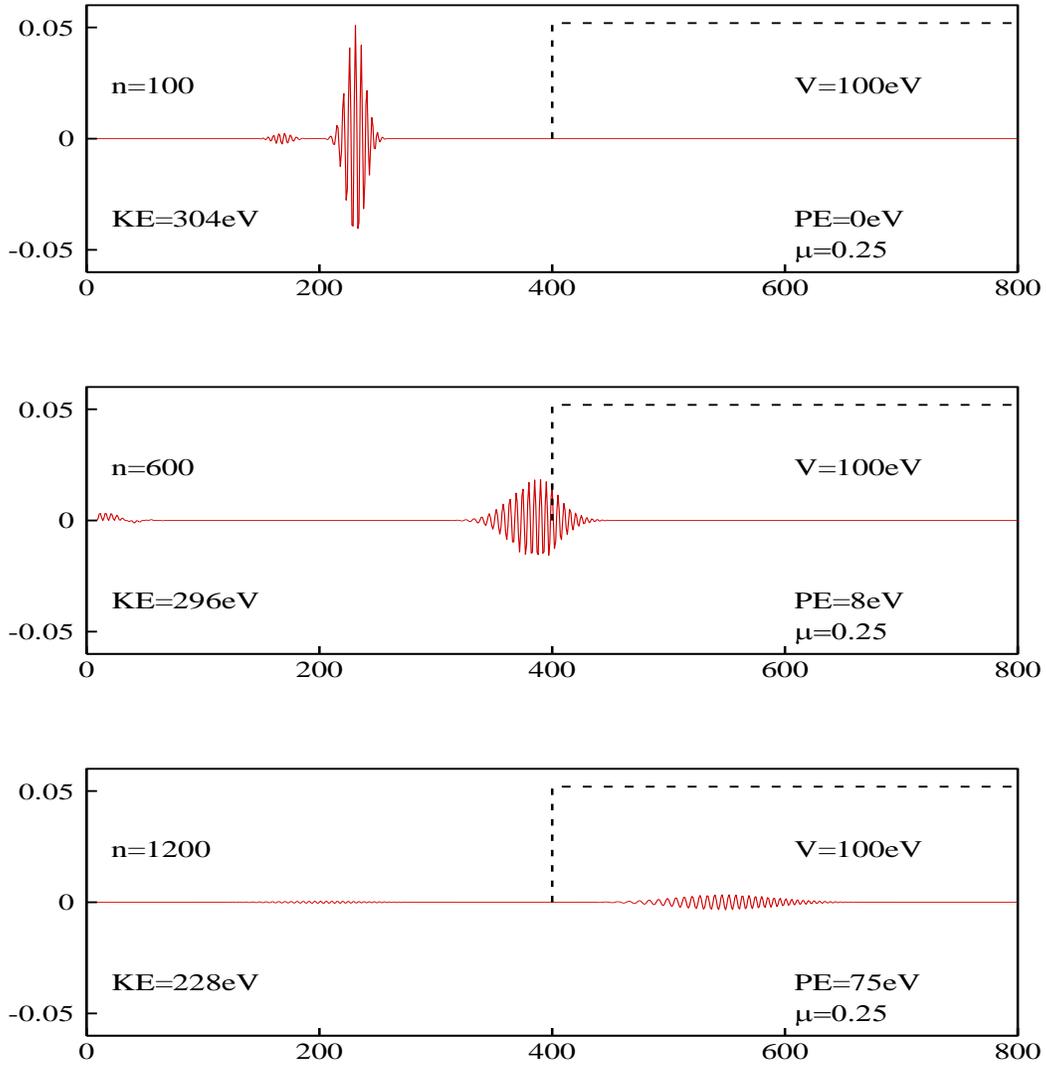

FIG. 5.  Simulation of an electron moving in 2-D free space and then hitting a potential. The generalized 2-D FDTD scheme, Eq. (38), was employed with $\mu = 0.25$.

14